# Breaking Walls: Pioneering Automatic Speech Recognition for Central Kurdish: End-to-End Transformer Paradigm

Abdulhady Abas Abdullah
Computer Science and Engineering Department
University of Kurdistan Hewlêr
Hewler, Kurdistan, Iraq
abdulhady.abas@ukh.edu.krd

Hadi Veisi
Faculty of New Sciences and Technologies
University of Tehran
Tehran, Iran
h.veisi@ut.ac.ir

Tarik Rashid
Computer Science and Engineering Department
University of Kurdistan Hewlêr
Hewler, Kurdistan, Iraq
tarik.ahmed@ukh.edu.krd

Abstract—End-to-end transformer-based models epitomize the cutting-edge in Automatic Speech Recognition (ASR) systems. Despite their substantial benefits, these models demand extensive training data to perform optimally, presenting a significant challenge for low-resource languages such as Central Kurdish. Addressing this issue requires innovative methods and techniques. This paper aims to develop an ASR system for Intermediate Kurdish by collecting a robust corpus of speech, using the N-GRAM language model, and utilizing an external Kurdish tokenizer for refinement and integration techniques to enhance the model's performance. We collect a comprehensive 100-hour speech corpus from diverse sources. Additionally, applied fine-tuning techniques to our speech corpus on Persian, English, and Arabic pre-trained models, specifically utilizing the xls-r-300m, xls-r-1b, and xls-r-2b Wav2vec 2.0 models. And utilized language models trained by 3-gram and 4-gram from a large text corpus of 300 million tokens. The fine-tuned xls-r-2b model, combined with a 3-gram language model and included external Kurdish tokenizer, achieved the best performance, yielding a Word Error Rate (WER) of 10.0% on the validation set and 11.8% on the Asosoft test set. The ASR model has demonstrated the advantages of having a large vocabulary compared to the existing Kurdish ASR models. Compared to other models, it produced more accurate and higher performance outcomes by working with a lower error rate.

## I. INTRODUCTION

Advances in automatic speech recognition (ASR) for high-resource languages such as English, German, and Russian have now surpassed human transcription accuracy (Li, J., 2022). In recent years, several technologies have emerged that leverage end-to-end deep learning models to enhance ASR performance. Notably, these models have outperformed traditional approaches in various languages and datasets, marking a significant leap forward in speech recognition technology. Despite these advancements, developing high-quality ASR systems for low-resource languages remains challenging due to the need for extensive and diverse labeled data. Nearly 7,000 languages worldwide lack access (Lewis et al., 2016) to the vast amounts of transcribed speech required to create efficient ASR systems. Kurdish, spoken by over 30 million people, is one such language. It is divided into three main dialects: Central Kurdish (CKB, i.e. Sorani), Northern Kurdish (Kurmanji), and Southern Kurdish. Although Kurdish has a large number of speakers, it faces significant difficulties in ASR development due to limited data resources. Several research efforts have aimed to address this issue. For instance, Veisi et al. (2022) presented Jira which is the first Large-vocabulary ASR system for Central Kurdish having used speech corpus and pronunciation dictionary to solve the problem of resource scarcity. The Jira Corpus with 43.68 hours long recordings from 576 speakers attained a word error rate of 13.9% on generic items by using SGMM model. One of the recent developments is the development of an ASR system for Middle Kurdish using Deep Neural Networks (DNN). This system employed Mel-frequency cepstral coefficients (MFCC), long-term short-term memory net- works (LSTM), and correlative temporal classification (CTC) that yielded a WER of 0.22%, significantly higher than previous CKB efforts (Abdullah and Veisi 2022). Nevertheless, significant obstacles persist despite these successes. Among others, existing ASR Corpuses such as Jira (Veisi et al., 2022) and LSTM ASR based (Abdullah and Veisi 2022) are not effective in real-time recognition in noisy environments; they have lower accuracy on children's voices while being sensitive to recording quality conditions. These limitations impede their usefulness for models to be utilized in everyday conversational settings and indicate the need for further progressions.

This paper aims to tackle these challenges by making several impactful contributions:

- Development of a Comprehensive Speech Corpus: We have curated a 100-hour speech corpus encompassing a diverse range of vocal variations, environmental sounds, and demographic characteristics, enhancing the representation and richness of the training data.

- Advancement of Self-Supervised Learning: We demonstrate the effectiveness of self-supervised learning techniques, specifically fine-tuning on pre-trained models such as xls-r-300m, xls-r-1b, and xls-r-2b Wav2vec 2.0, tailored for the CKB dialect to improve ASR performance in low-resource settings.
- Integration of Language Models: We incorporate n-gram language models (3-gram and 4-gram) trained on a large text corpus of 300 million tokens, which contributes to enhanced accuracy in word recognition and transcription.
- Conducting extensive experiments and analyses to uncover significant insights and improvements in Kurdish ASR.

The structure of this paper is as follows: In Section 2, the related works with speech recognition technology are described. Section 3 delves into the creation and design of the speech corpus specifically tailored for the CKB language. The proposed method and the architecture of the applied model for the CKB language ASR system are explained in Section 4. In Section 5, the results and discussions are presented. This paper is then concluded in Section 6.

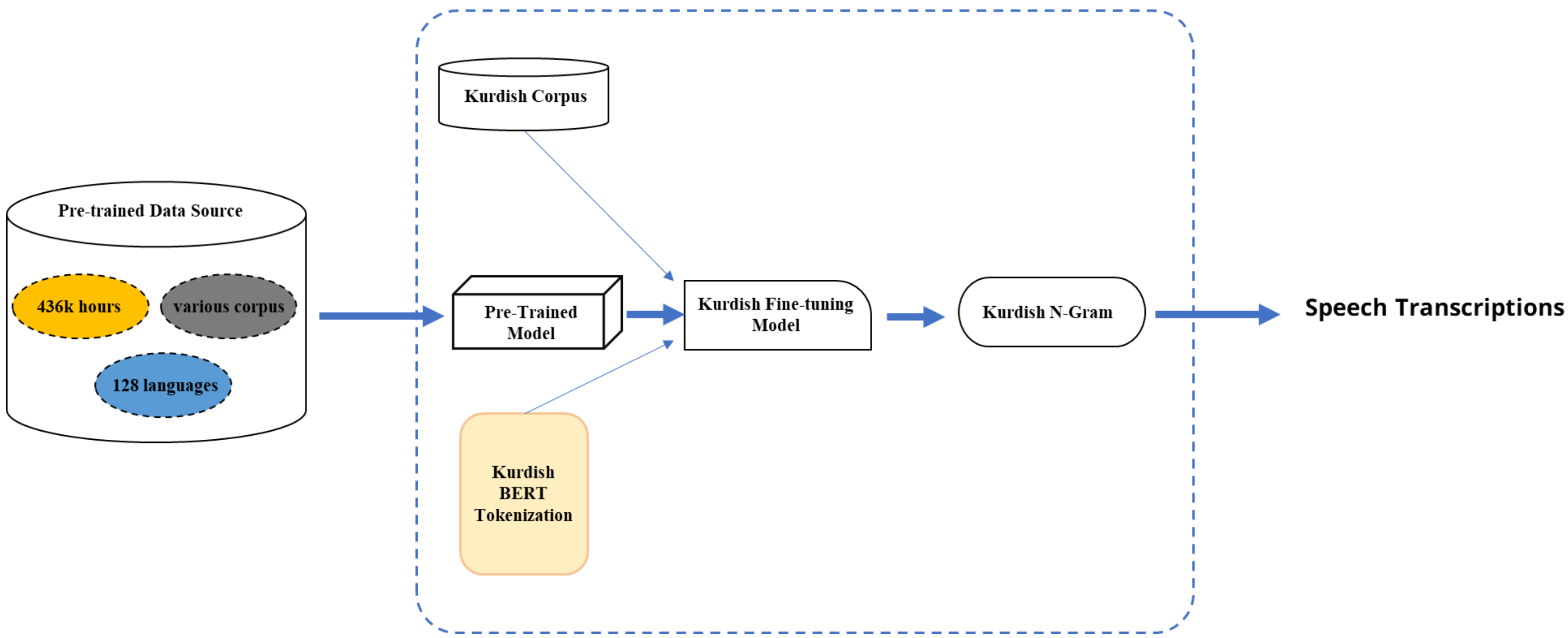

Fig. 1. The Main Architecture Method Proposed

## II. Proposed Method

This section describes the proposed methodology for developing a Kurdish ASR model using a pre-trained multilingual model and a fine-tuning process specifically tailored for the Kurdish Corpus. The process involves several key steps, utilizing Kurdish speech corpus, model fine-tuning, and N-gram integration. The overall architecture is illustrated in the provided Fig. 1.

### A. *Kurdish Speech Corpus*

This section describes the steps followed in Fig. 1 to establish and gather a new speech corpus for the CKB language that is mainly meant to improve ASR in various domains and environments. The main aim is to increase the speech corpus sources for a target language as well as mitigate limitations found in the existing corpus, such as AsoSoft (Veisi et al. 2022) and Mozilla (Ardila et al., 2019).

*a) Designing and Recording Sentences:* Speech corpus creation involves collecting and curating sentences, recording them with standardized methods, and then conducting manual checks for accuracy. Data is subsequently normalized to ensure consistency and quality, as illustrated in Fig. 2.

First, Corpus sentence collection needs to cover the acoustic variations of Kurdish as a process requirement. In this study designed and generated about 1,000 sentences to cover the entire Central Kurdish language consisting of a variety of sentences in different genres. The statistics of these sentences are mentioned in Table 1[1].

[1] https://github.com/AsoSoft/CKB-Sentence-Dataset

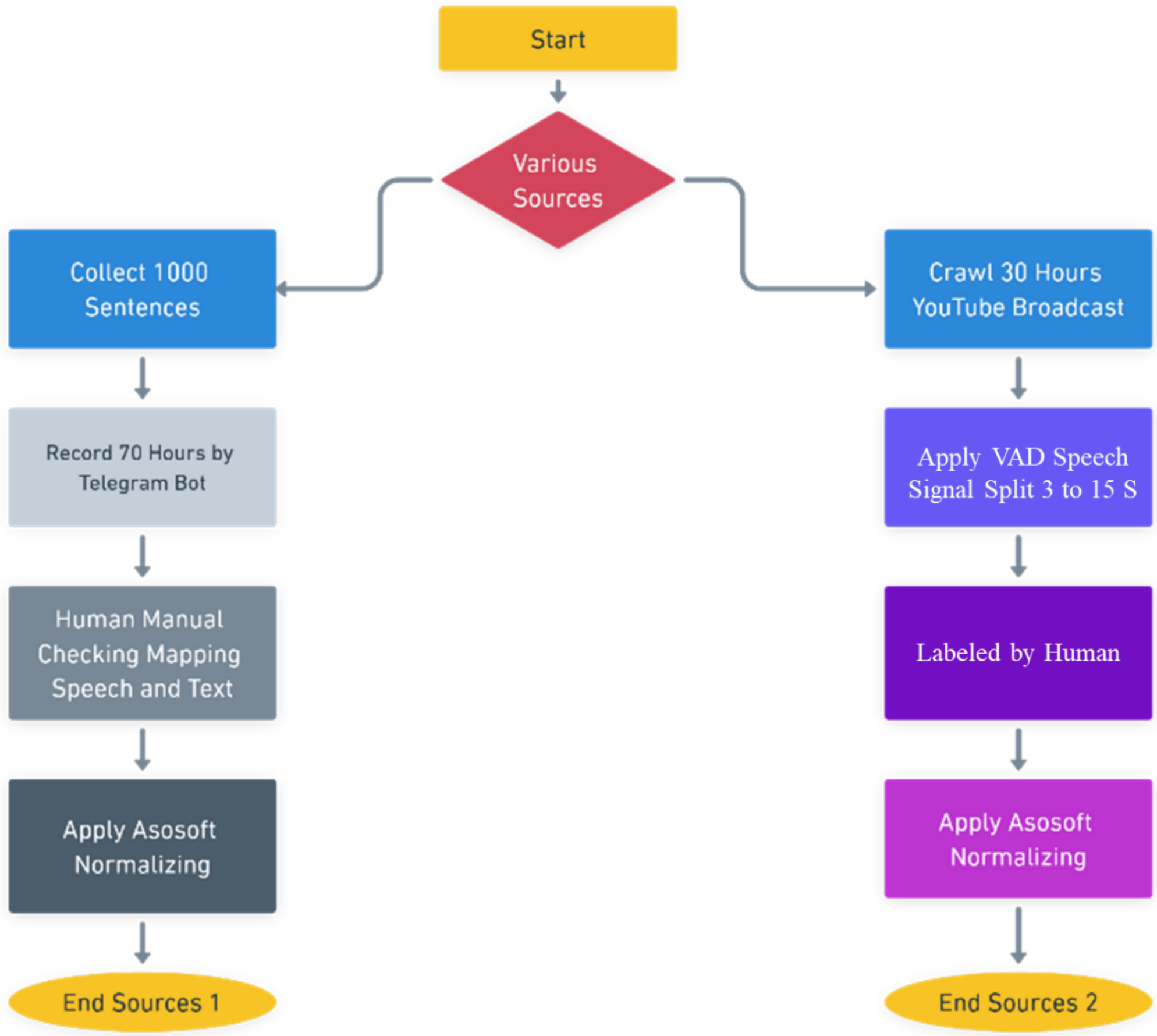

Fig. 2. Workflow for Speech Corpus Development and Normalization

TABLE 1: TOTAL CKB SENTENCES CORPUS.

| Topics | Number of sentences |
|---|---|
| **History** | 50 |
| **Geography** | 50 |
| **Sports** | 100 |
| **Religious** | 50 |
| **General** | 50 |
| **News** | 50 |
| **Health** | 50 |
| **Weather** | 50 |
| **Arts** | 50 |
| **Science and Technology** | 50 |
| **Poetry** | 100 |
| **Economy** | 50 |
| **Very common** | 50 |
| **Facebook comment** | 50 |
| **Government** | 50 |
| **Normal** | 100 |
| **Total** | **1000** |

Some of the sentences designed to create a speech corpus for CKB are listed in Table 2.

TABLE 2: SOME SAMPLES OF THE DESIGNED SENTENCES IN THE CKB SPEECH DATASET.

| Title of sentences | Kurdish | English |
|---|---|---|
| **Sport** | بۆ ئامادەبوون لە یاریگەدا و بینینی یاریی سعودیە و ژاپۆن. | to attend the stadium and see Saudi and Japanese games. |
| **Health** | دکتۆرەکان داوای ئۆکسجینی زیاتر ئەکەن بۆ نەخۆشیەکانی کۆرۆنا | Doctors request more oxygen for corona diseases |
| **News** | تەنیا کارە پرۆفیشناڵەکان لێرە نمایش دەکرێن | Only professional works are shown here. |
| **History** | کوردەکان نەتەوەیەکی نیشتەجێی ڕۆژهەڵاتی ناوینن. | The Kurds are a nation residing in the Middle East. |

Second, create a Kurdish Telegram bot[2], which facilitates the collection of speech data by interacting with users via Telegram. It sends predefined sentences to users, who then

[2] https://t.me/STTkurdish_bot

record their own speech reading those sentences, as shown interface bot in Fig. 3.

*Key Features:*

- *Sending Sentences:* The bot sends text sentences to users for recording.
- *Voice Recording*: Users respond with voice messages containing their recordings.
- *Processing:* The bot processes and formats the recordings, ensuring quality and usability.
- *Data Management:* Recordings are stored with metadata for easy retrieval.
- *Feedback:* The bot provides feedback to users and sends reminders to encourage completion.

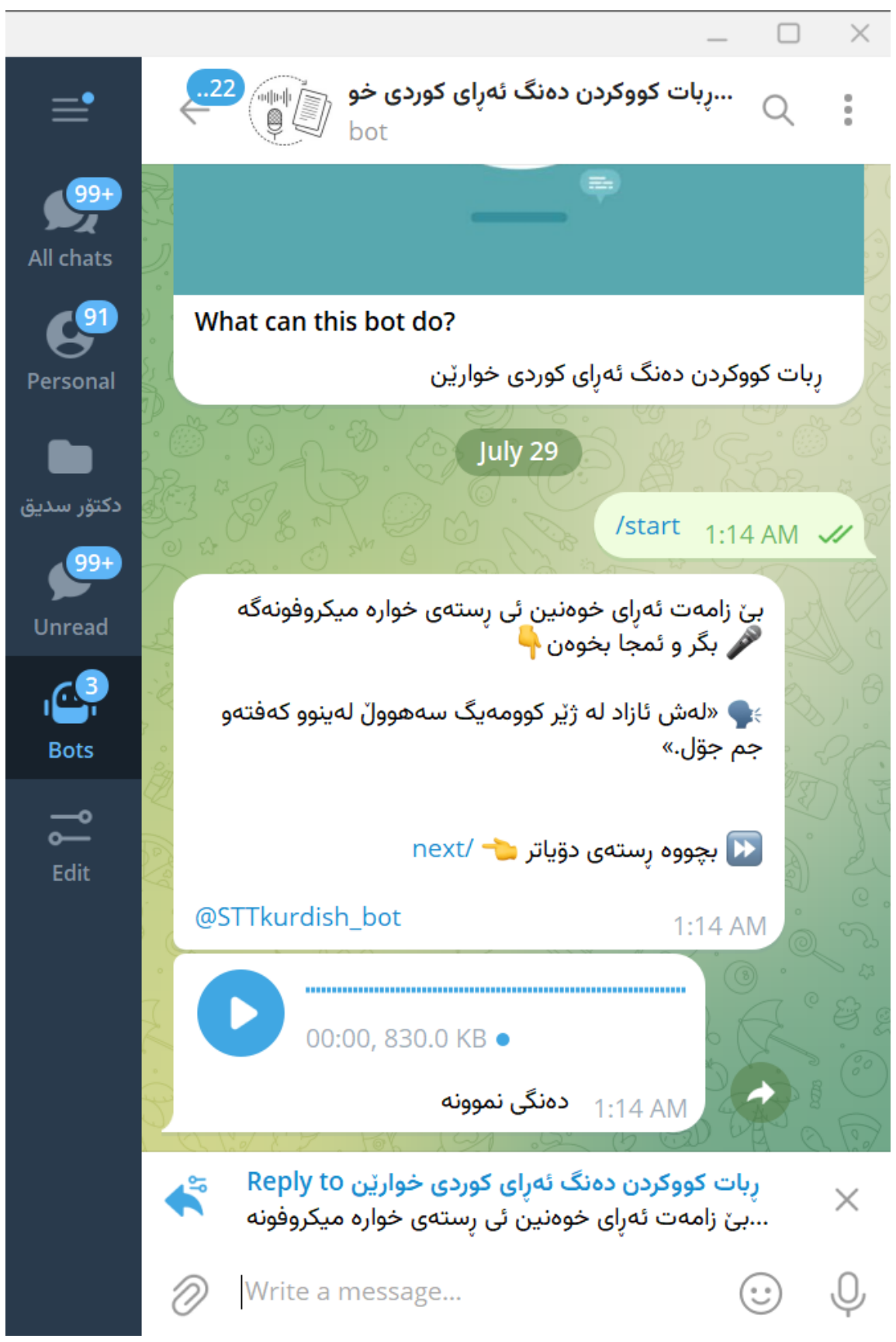

Fig. 3. Kurdish Telegram Bot for Recording Sentences

Third, Fig. 4 shown web app for manual verification of sentence accuracy involves reviewing the recorded speech against the provided text sentences. Users can listen to the audio recordings and compare them with the original text to ensure that the spoken content accurately reflects the written sentence. This process helps identify discrepancies, such as mispronunciations or incomplete recordings, ensuring high-quality data for ASR systems.

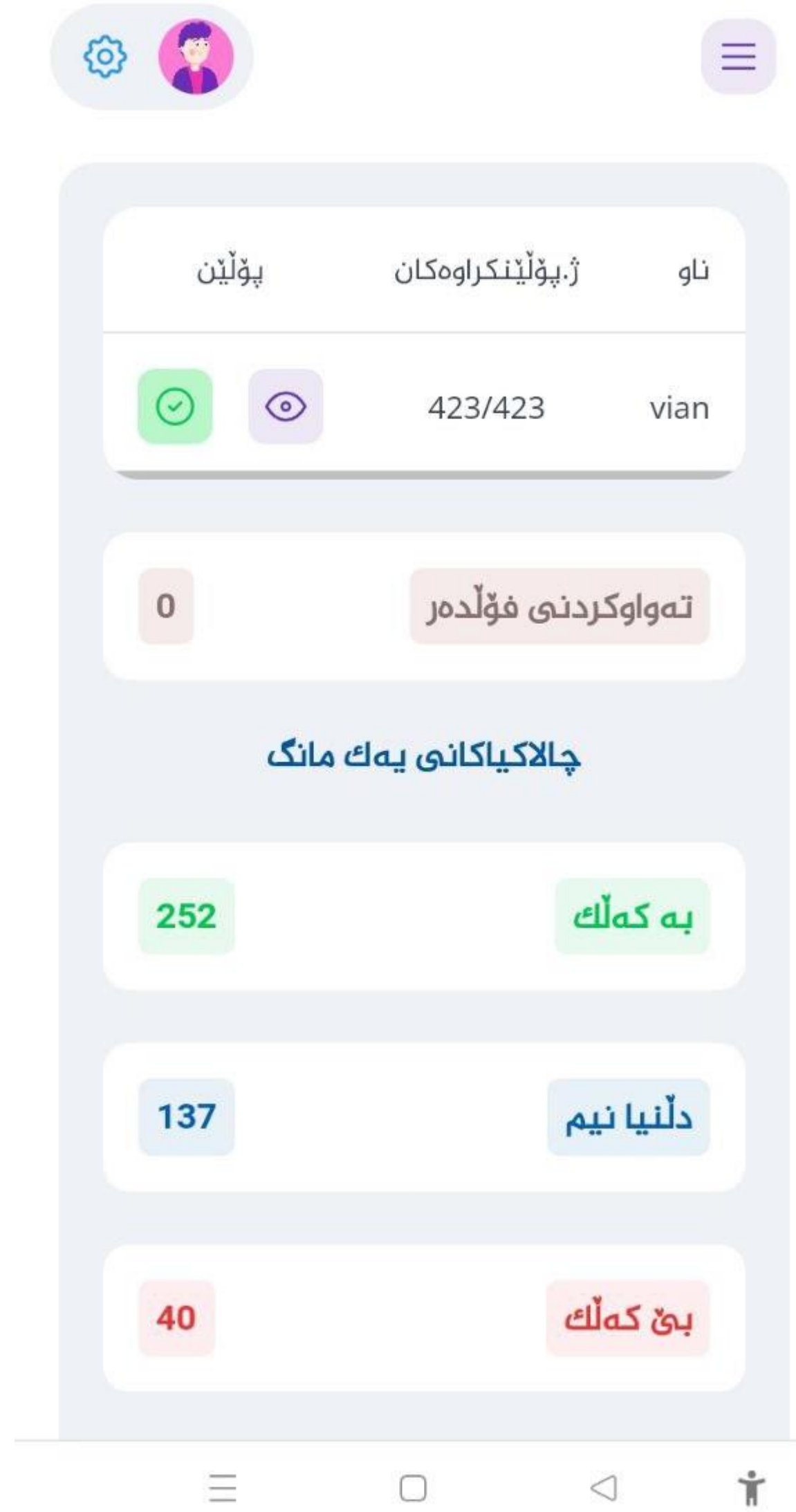

Fig. 4. Web App Platform for Manual Verification Corpus

Finally, Table 3 below provides an overview of the dataset used for the project, including the total and validated duration of recordings, the number of speech files, and the distribution of speakers' genders. The data was collected from a Telegram crawl corpus, totaling 95 hours of recordings with 1,416 speech files and 543,000 speech samples, Table 4 shown metadata speech corpus.

TABLE 3: SUMMERED RECORDED SPEECH DATA

| Attribute | Details |
|---|---|
| **Total Dataset Duration** | 95 hours |
| **Validated Duration** | 70 hours |
| **Number of Speech Files** | 1,416 |
| **Total Speech Samples** | 543,000 |
| **Gender Distribution** | 69% male, 31% female |
| **Source** | Telegram crawl corpus |

TABLE 4: METADATA OF SPEECH CORPUS

| mp3_filename | mp3_filesize | transcript |
|---|---|---|
| **001MLU606.mp3** | 152962 | هەست بە کێشە و ئازارەکانی خەڵک ئەکەم |
| **001MLU609.mp3** | 170882 | خەونەکان پێش ئەوە بێنە دی لە سێدارە دران |
| **001MLU608.mp3** | 148802 | لە کۆڕێکدا تێرم باسی بیروڕای خۆم کرد |

*b) Crawl YouTube Broadcast*: First, we started by collecting playlists that included children's voices, conversations, and various audio recordings from studio environments and diverse settings, Table 5 some examples playlists utilized.

TABLE 5: EXAMPLE OF PLAYLIST USED

| YouTube Channel | Playlist Name | Number of Videos | Total Time |
|---|---|---|---|
| **Afarin Kids TV** | بەرهەمەکانی هودهود | 54 | 02:10:50 |
| **Rudaw TV** | باوەڕنما | 55 | 05:43:28 |
| **Ava TV** | Mallî Kurdo | 7 | 08:24:54 |

As a result, 100 hours of Kurdish[3] speech labeled data were obtained from both sources after cleaning and normalization by using the AsoSoft Library[4] (Mohammadamini et al., 2019), summarized in Table 6.

TABLE 6: GENERAL SPEECH CROUPS

| Attribute | Details | Total Speaker | Gender |
|---|---|---|---|
| **Total Sentence record Duration** | 70 hours | 1,416 | 69% male, 31% female |
| **Total YouTube labeled Duration** | 30 hours | 55 | 55% male, 45% female |
| **Total** | 100 hours | 1471 | 62% male, 38% female |

Finally, we wrote a script to convert these playlists into audio files and segmented them into smaller chunks ranging from 3 to 15 seconds by using Voice activity detection (VAD). Afterward, the data was manually labeled by utilizing a web app platform for labeling shown in Fig. 4.

### B. *Wav2Vec Pretrained Models*

XLS-R is a scalable model for learning speech representation languages based on the Wav2Vec 2.0 framework shown in Fig. 6. The XLS-R series includes variable parameter models, specifically the XLS-R-300M, XLS-R -1B, and XLS-R-2B, trained on 436,000 hours of totally available speech sounds in 128 languages, summarized in Table 7. Functions speech processing. XLS-R models use self-supervised learning techniques to encode raw sound into meaningful representations, which are then elaborated into specific tasks such as automatic speech recognition (ASR), and speech translation. Its performance on measures such as CoVoST-2, BABEL, MLS, CommonVoice, and VoxPopuli shows significant improvements, reducing error rates by 20% to 33% compared to average and setting new standards in speech processing.

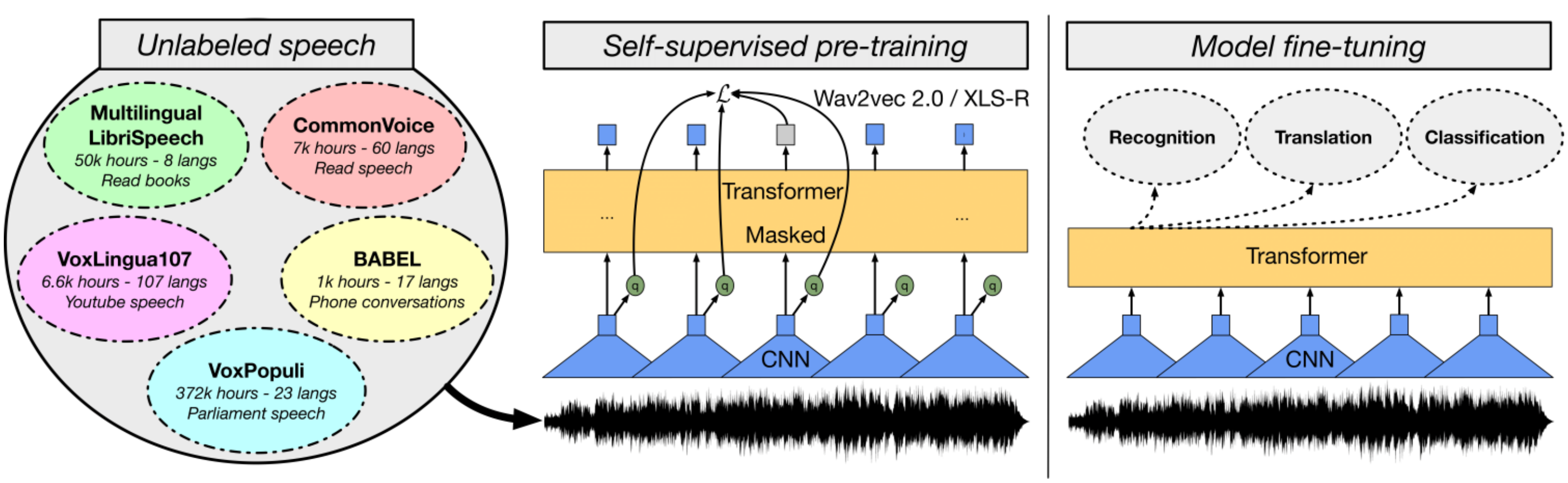

Fig. 5. wav2vec2.0 Pre-Training model (Baevski et al., 2020)

3 https://github.com/AsoSoft/CKB-New-Speech-Corpus

4 https://github.com/AsoSoft/AsoSoft-Library

TABLE 7: WAV2VEC 2.0 XLS-R MODELS

| Model | Parameters | Total Hours | N Languages |
|---|---|---|---|
| XLS-R-300M | 300 million | 436,000 | 128 |
| **XLS-R-1B** | 1 billion | 436,000 | 128 |
| **XLS-R-2B** | 2 billion | 436,000 | 128 |

### *C. Fine-Tuning*

The fine-tuning of Wav2Vec 2.0 is particularly important for Kurdish low resource language, Kurdish faces significant challenges due to limited amount of data. Fine-tuning goal for improving quality and reduce error rate. Although Wav2Vec 2.0 was pre-trained on large datasets, this training mainly includes 128 high resource languages, unfortunately Kurdish is not including. In this work we will focus on the process of fine-tuning Kurdish speech corpus by using various pre-train Wav2Vec 2.0 model. Fine-tuning in ASR involves adapting a pre-trained model to a specific target language to enhance its accuracy. This process uses high-level acoustic features learned by pre-trained models on large corpus, and transfers these learned features to the target language, which may be underrepresented and low resources. By fine-tuning, the model adjusts its parameters based on the specific phonetic and linguistic features of the target language, improving the ability to accurately recognize and transcribe speech. fine-tuning allows models to effectively transfer common acoustic patterns and features acquired from high source language to unique speech patterns of the target language, resulting in better recognition performance and greater accuracy, fine-tuning substituting large amounts of data for low resource Languages.

### *D. N-gram Language Model*

The integration of N-gram models into the decoder part of the architecture ASR model is a way to improve the accuracy and reduce the error rate, especially in these languages with a small amount of data to which. The N-gram language model is a statistical approach used to predict the order probability of words based on previous words present on a large portable text of the target language. An n-gram model array, which determines how many previous words to use to predict the next word. Common options are bigram (2-gram) and three-gram (3-gram) for languages that do not have the necessary amount of data, for languages with a large amount of data, more (4-gram) and (5-gram) are used, which are shown in Fig. 6. As well as, in short, an N gram language model is a list of n words. The model is therefore used to suggest that the following words rely on a particular textual corpus. Since the n-gram model is based on probability, it is possible (Khandelwal et al., 2018). Consequently, previous work is used in the expected (n-1) Markov chain model structure. The simplicity and scalability of n-gram models containing n or more numbers or values are its main advantages.

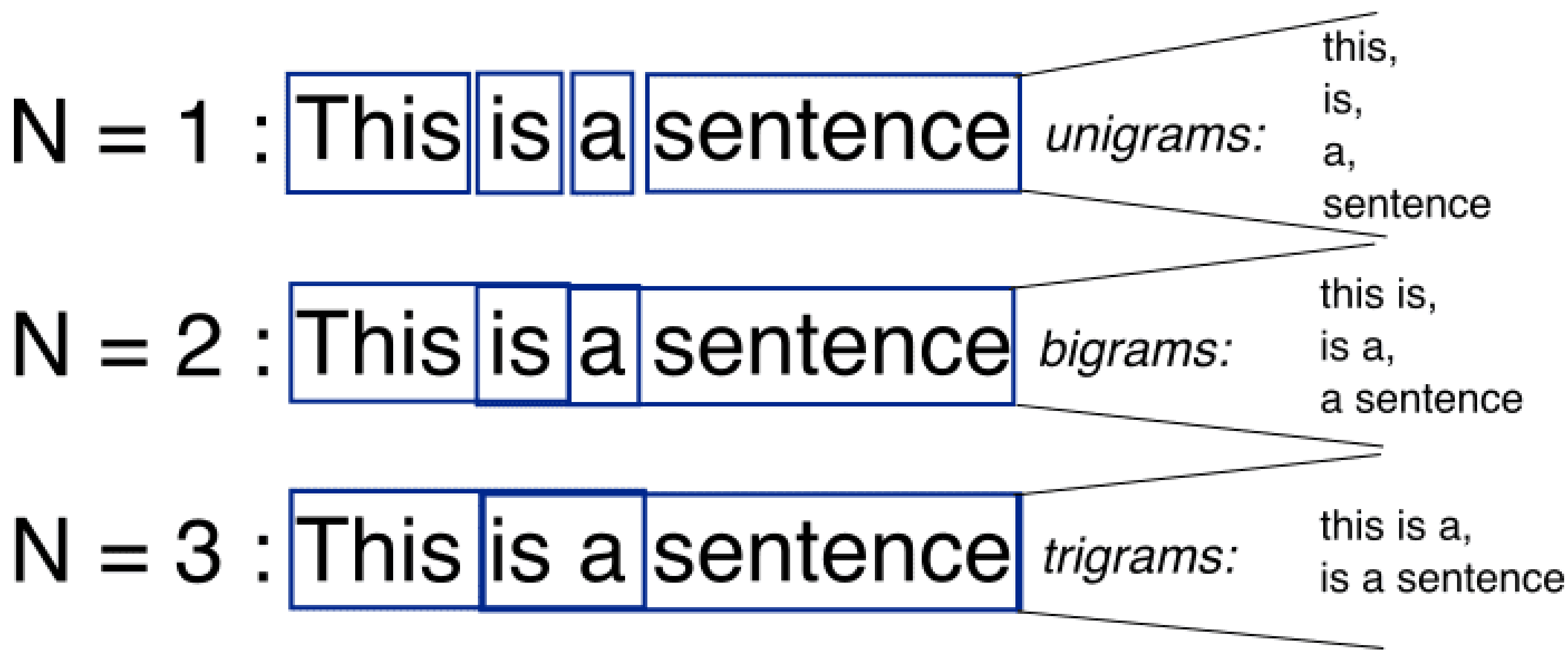

Fig. 6. N-gram Language Model Integration in Decoder Network

## III. EXPERIMENTS

In this section, we describe the experimental setup including the speech corpuses, model setups and implementations, evaluation metric and the experimental results, and the analysis of the experiments conducted to evaluate the performance of the proposed ASR system for CKB. We mainly evaluate the ability of fine-tuning the ASR model on our corpus, how bigram language models help the performance and including external Kurdish tokenizer for it.

### *A. ASR Evaluation Criteria*

ASR systems' performance is usually measured in terms of accuracy and speed. Accuracy is generally measured in terms of performance accuracy, usually expressed as a word error rate (WER), whereas speed is described as a real-time factor. Single Word Error Rate (SWER) and Command Error Rate (CSR) are two other accuracy measures. Word Error Rate (WER) and Word Recognition Rate (WRR) are used to evaluate the speech recognizer's performance (Wigington et al., 2017). Several types of word errors include insertions, substitutions, and deletions.

Finally, the following equations are used to calculate the word error rate and word recognition rate.

Word Error Rate(%) =(Insertion(I)+Substitution(S)+Deletion(D))/(No.of Reference Words(N))*100 (1)

Word Recognition Rate (WRR)=1-WER =(N-S-D-I)/N (2)

Character Error Rate (CER) is another popular measure of ASR system performance. CER is similar to WER but acts on the character instead of the word.

*B. Speech Dataset*

We used a 100-hour speech corpus generated from this work for training and validation, and the external test speech suite AsoSoft by Veisi et al. (2022) to check the models. The AsoSoft test set[5] comprises 1.25 hours of speech from 100 sentences related to various fields recorded by eight speakers in an office setting and a further 50 minutes of speech from 135 speakers in crowded source settings and from various backgrounds. It is such a test suite that offers diverse acoustic conditions, from the controlled environments of offices to various crowd-sourced contexts.

*C. Model Configurations and Wav2Vec 2.0 Fine-Tuning*

This section, focuses on describing the data and hyperparameters used to fine-tune our automatic speech recognition (ASR) model for Kurdish. Specifically, we fine-tuned pre-trained Wav2Vec 2.0 models (XLS-R-300M, XLS-R-1B, and XLS-R-2B) that were initially trained on large-scale Persian, English, and Arabic datasets. These models were retrained with 100 hours of semi-manually transcribed Kurdish speech data. A 90% to 10% split of the corpus was made between training and validation to maintain a balanced train-validation strategy. A high-performance computing environment using 48 NVIDIA RTX 4090 GPUs was employed in the UKH AI Center lab to carry out the fine-tuning. This powerful infrastructure facilitated efficient processing and optimization. Key hyperparameters were fine-tuned to enhance the model's robustness and efficiency. Attention Dropout, Hidden Dropout, and Feature Projection Dropout (feat_proj_dropout) were set to 0.1 across layers to prevent overfitting. The Layer Drop parameter was set to 1, allowing entire layers to be dropped during training, increasing the model's robustness. Gradient checkpointing was used to manage memory usage efficiently during backpropagation, enabling the model to be trained on limited hardware. The training batch size per device was set to 32, indicating the number of examples processed in each iteration. The training process involved 35,000 steps, allowing the model to traverse the entire dataset multiple times, improving its understanding and performance. Additionally, external Kurdish tokenization was used to efficiently split the text into manageable pieces, or tokens. We employed Kurdish BERT Tokenization (Veisi et al., 2024), trained on a dataset of around 300 million tokens[6]. Furthermore, 3-gram and 4-gram language models were trained on the same corpus for benchmarking (Abdullah & Veisi, 2022).

*D. Results*

In The section describes the experimental design used for testing ASR systems, focusing on fine-tuning three pre-trained Wav2Vec 2.0 models (XLS-R-300M, XLS-R-1B, and XLS-R-2B) on a new corpus of Kurdish speech. For each of the models, we report the results of testing with different languages for pretraining: Arabic, Persian, English, and components presented both without and with an external Kurdish tokenizer integration. In further improving the output layer accuracy, the language models were integrated into the ASR network decoder for the 3-gram and 4-gram. As with the XLS-R-300M model, one could observe that performance had generally improved under the introduction of the external Kurdish tokenizer in all language pre-training types, inclusive of the 3-gram Kurdish language model. As a result, utilized Persian pre-training with an external tokenizer and integration 3-gram language model resulted in a validation WER of 12.0% and an Asosoft test WER of 14.8% reported. This alignment consistently outperformed the other components, indicating the beneficial role of specialized marking and specific language modeling, the results are summarized in Table 8.

Subsequently, the XLS-R-1B model followed a similar experimental approach. utilized pre-training Persian language again showed the most significant improvement when an external Kurdish tokenizer and LM of 3 grams were used, obtaining the best efficiency measures with validation WER of 11.1% and Asosoft test WER of 12.9%, the results are summarized in Table 9.

Lastly, for model XLS-R-2B, the pre-training Persian language component, with a Kurdish external tokenizer and LM of 3 grams, obtained the lowest WER score in the validation and test sets. The model performance peaked with a validation WER of 10.0% and the AsoSoft test set WER of 11.8%, the results are summarized in Table 10. Accordingly, Table 11 presents the benchmark WER results for Kurdish speech recognition using various XLS-R models, ranging from 300 million to 2 billion parameters. The models were pre-trained on multiple languages and used an external tokenizer as well as a 3-gram language model for decoding. The validation WER and test set WER on the AsoSoft dataset show a decreasing trend as the model size increases. Specifically, the xls-r-300m model achieved a validation WER of 12.0% and the test set a WER of 14.8%. The xls-r-1b model improved performance, reducing the validation WER to 11.1% and the test set the WER to 12.9%. The largest model, xls-r-2b, showed the best results with a validation WER of 10.0% and a test set WER of 11.8%. These results indicate that increasing the model size from 300 million to 2 billion parameters systematically improves model performance, leading to a decline in WER. The consistent use of an external tokenizer and a 3-gram linguistic model suggests a robust approach to decoding across all trials. Consequently, compared to other Kurdish ASR models as in the AsoSoft test set illustrated in Fig. 7, the xls-r-2b with the integration of the 3-gram external Kurdish tokenizer recorded a lower WER. For example, the LSTM + 3-gram model, as studied in papers (Abdullah and Veisi, 2022), recorded a WER of 22.0, while the Jira model scored exactly in another study (Veisi et al., 2022), WER of 22.0 25. These significant differences in WER highlight the advanced performance of the CKB-XLS-R-2B model, positioning it as a leader in Kurdish ASR technology development.

---

5 https://github.com/AsoSoft/AsoSoft-Speech-Corpus

6 https://github.com/AsoSoft/KuBERT-Central-Kurdish-BERT-Model

TABLE 8: XLS-R-300M MODEL RESULTS

| Pre-Train Model | Languages | External Tokenizer | LM | Validation (WER) | Asosoft Testset (WER) |
|---|---|---|---|---|---|
| xls-r-300m | Arabic | Included | 3-gram | 13.2 | 15.3 |
| **xls-r-300m** | **Persian** | **Included** | **3-gram** | **12** | **14.8** |
| xls-r-300m | English | Included | 3-gram | 13.7 | 15.8 |

TABLE 9: XLS-R-1B MODEL RESULTS

| Pre-Train Model | Languages | External Tokenizer | LM | Validation (WER) | Asosoft Testset (WER) |
|---|---|---|---|---|---|
| xls-r-1b | Arabic | Included | 3-gram | 11.8 | 13.2 |
| **xls-r-1b** | **Persian** | **Included** | **3-gram** | **11.1** | **12.9** |
| xls-r-1b | English | Included | 3-gram | 12.6 | 13.6 |

TABLE 10: XLS-R-2B MODEL RESULTS

| Pre-Train Model | Languages | External Tokenizer | LM | Validation (WER) | Asosoft Testset (WER) |
|---|---|---|---|---|---|
| xls-r-2b | Arabic | Included | 3-gram | 11.5 | 12.8 |
| **xls-r-2b** | **Persian** | **Included** | **3-gram** | **10.0** | **11.8** |
| xls-r-2b | English | Included | 3-gram | 12.2 | 13.3 |

TABLE 11: BENCHMARK RESULTS FOR WER USING XLS-R MODELS

| Pre-Train Model | Languages | External Tokenizer | LM | Validation (WER) | Asosoft Testset (WER) |
|---|---|---|---|---|---|
| **xls-r-300m** | Persian | Included | 3-gram | 12.0 | 14.8 |
| **xls-r-1b** | Persian | Included | 3-gram | 11.1 | 12.9 |
| **xls-r-2b** | **Persian** | **Included** | **3-gram** | **10.0** | **11.8** |

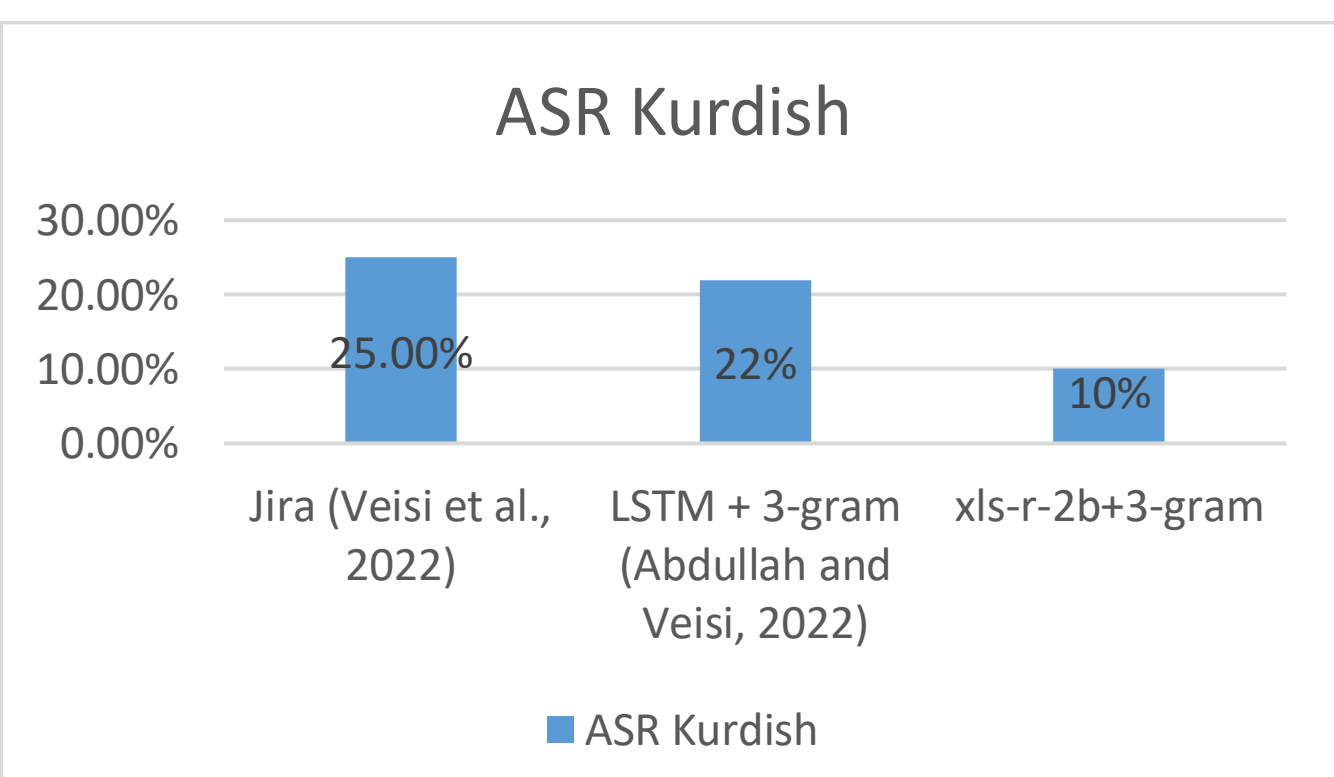

Fig. 7. Benchmark WER for ASR Kurdish models

### *E. Discussion*

This section highlights several factors that are major contributors to achieving high accuracy and quality in the Kurdish ASR model. The impact results obtained utilized pre-train Persian have been compiled using a 3-gram language model and an integration external Kurdish tokenizer. The superior performance of XLS-R-2B Persian pretraining can be attributed to the fact that the linguistic similarities between Persian and Kurdish, such as close vocabulary and phonological characteristics, which contributed to impact results compared to Arabic and English. Additionally, the 3-gram language model outperformed the 4-gram model are due to the limited size of the corpus used, which totaled 300 million tokens. This quantity is insufficient to develop robust 4-gram model, hence the 3-gram model delivered the best performed. Moreover, the utilized of external Kurdish tokenizers significantly enhancing accuracy across various cases, as these tokenizers improving the model's ability to learn Kurdish data effectively during the tokenization process of speech dataset sentences.

## IV. CONCLUSIONS

in this study focus development of an advanced automatic speech recognition (ASR) system for Central Kurdish, a low-resource language, using end-to-end transformer-based models. Our contributions are as follows: 1) a 100-hour comprehensive speech corpus has been developed; 2) pre-training of fine-tuned Wav2Vec 2.0 models (XLS-R-300M, XLS-R-1B, and XLS-R-2B) on this corpus; and 3) integration of N-gram language models (3-gram and 4-gram) for enhancing performance. our approach showed the successful use of pre-trained multilingual models, specifically fine-tuning them with a comprehensive corpus of Kurdish. By integrating an external Kurdish tokenizer and a 3-gram language model into the models, we significantly improved performance. In our experiments, the XLS-R-2B model that had been fine-tuned with Persian pre-trained, an external tokenizer and a 3-gram language model achieved the best results with a validation WER of 10.0% and an Asosoft test set WER of 11.8%. In conclusion, this study has made a significant step forward in achieving a practical and accurate ASR system for CKB. Moreover, the results highlight the potential of advanced transformer models and tailored fine-tuning strategies for enhancing the ASR technologies of low-resource languages.

## REFERENCES

[1] Abdullah, A.A. and Veisi, H., 2022. Central Kurdish Automatic Speech Recognition using Deep Learning. Journal of university of Anbar for Pure science, 16(2).

[2] Li, J., 2022. Recent advances in end-to-end automatic speech recognition. APSIPA Transactions on Signal and Information Processing, 11(1).

[3] Ardila, R., Branson, M., Davis, K., Henretty, M., Kohler, M., Meyer, J., Morais, R., Saunders, L., Tyers, F.M. and Weber, G., 2019. Common voice: A massively-multilingual speech corpus. arXiv preprint arXiv:1912.06670.

[4] Baevski, A., Zhou, Y., Mohamed, A. And Auli, M., 2020. Wav2vec 2.0: A Framework For Self-Supervised Learning Of Speech Representations. Advances In Neural Information Processing Systems, 33, Pp.12449-12460.

[5] Gupta, V. And Boulianne, G., Progress ,2022, In Multilingual Speech Recognition For Low Resource Languages Kurmanji Kurdish, Cree And Inuktut.

[6] Hagen, A. and Morris, A., 2005. Recent advances in the multi-stream HMM/ANN hybrid approach to noise robust ASR. Computer Speech & Language, 19(1), pp.3-30.

[7] Imaizumi, R., Masumura, R., Shiota, S. and Kiya, H., 2020, December. Dialect-aware modeling for end-to-end Japanese dialect speech recognition. In 2020 Asia-Pacific Signal and Information Processing Association Annual Summit and Conference (APSIPA ASC) (pp. 297-301). IEEE.

[8] Kim, J. And Kang, P., 2021. K-Wav2vec 2.0: Automatic Speech Recognition Based On Joint Decoding Of Graphemes And Syllables. Arxiv Preprint Arxiv:2110.05172.

[9] Laptev, A., Andrusenko, A., Podluzhny, I., Mitrofanov, A., Medennikov, I. And Matveev, Y., 2021. Dynamic Acoustic Unit Augmentation With Bpe-Dropout For Low-Resource End-To-End Speech Recognition. Sensors, 21(9), P.3063.

[10] Lewis, M. P., Simon, G. F., & Fennig, C. D. (2016). Ethnologue: Languages of the world (nineteenth edition). Retrieved from http://www.ethnologue.com

[11] Muhamad, S. and Veisi, H., 2022. End-to-End Kurdish Speech Synthesis Based on Transfer Learning. Passer Journal of Basic and Applied Sciences, 4(2), pp.150-160.

[12] Vaswani, A., Shazeer, N., Parmar, N., Uszkoreit, J., Jones, L., Gomez, A.N., Kaiser, Ł. And Polosukhin, I., 2017. Attention Is All You Need. Advances In Neural Information Processing Systems, 30.

[13] Veisi, H., Hosseini, H., Mohammadamini, M., Fathy, W. And Mahmudi, A., 2022. Jira: A Central Kurdish Speech Recognition System, Designing And Building Speech Corpus And Pronunciation Lexicon. Language Resources And Evaluation, Pp.1-25.

[14] Veisi, H., MohammadAmini, M. and Hosseini, H., 2020. Toward Kurdish language processing: Experiments in collecting and processing the AsoSoft text corpus. Digital Scholarship in the Humanities, 35(1), pp.176-193.

[15] Wang, C., Tang, Y., Ma, X., Wu, A., Okhonko, D. and Pino, J., 2020. fairseq s2t: Fast speech-to-text modeling with fairseq. arXiv preprint arXiv:2010.05171.

[16] Wigington, C., Stewart, S., Davis, B., Barrett, B., Price, B. and Cohen, S., 2017, November. Data augmentation for recognition of handwritten words and lines using a CNN-LSTM network. In 2017 14th IAPR international conference on document analysis and recognition (ICDAR) (Vol. 1, pp. 639-645). IEEE.

[17] Veisi, H., muhealddin Awlla, K. and Abdullah, A.A., 2024. KuBERT: Central Kurdish BERT Model and Its Application for Sentiment Analysis.

[18] Zhou, S., Dong, L., Xu, S. And Xu, B., 2018. Syllable-Based Sequence-To-Sequence Speech Recognition With The Transformer In Mandarin Chinese. Arxiv Preprint Arxiv:1804.10752.

[19] Zouhair, T., 2021. Automatic Speech Recognition For Low-Resource Languages Using Wav2vec2: Modern Standard Arabic (Msa) As An Example Of A Low-Resource Language.